\documentclass[reprint]{revtex4-1}
\usepackage{graphicx}
\usepackage{amssymb}
\usepackage{amsmath}
\usepackage{epsfig}
\usepackage{chemarr}
\usepackage{verbatim}
\usepackage{url}
\usepackage{natbib}
\usepackage[final]{changes}
\usepackage{dcolumn}
\usepackage{bm}
\usepackage{color}

\def\beq{\begin{equation}}
\def\eeq{\end{equation}}

\def\beqa{\begin{eqnarray}}
\def\eeqa{\end{eqnarray}}

\begin{document}
\title{Associative pattern recognition through macro-molecular self-assembly}
\author{Weishun Zhong$^1$, David J. Schwab$^2$, Arvind Murugan$^1$}
\affiliation{${}^1$ Department of Physics and the James Franck Institute, University of Chicago, Chicago, IL 60637 \\
${}^2$ Department of Physics, Northwestern, Evanston, IL 60000} 

\begin{abstract}
We show that macro-molecular self-assembly can recognize and classify high-dimensional patterns in the concentrations of $N$ distinct molecular species. Similar to associative neural networks, the recognition here leverages dynamical attractors to recognize and reconstruct partially corrupted patterns. Traditional parameters of pattern recognition theory, such as sparsity, fidelity, and capacity are related to physical parameters, such as nucleation barriers, interaction range, and non-equilibrium assembly forces. Notably, we find that self-assembly bears greater similarity to continuous attractor neural networks, such as place cell networks that store spatial memories, rather than discrete memory networks.  
This relationship suggests that features and trade-offs seen here are not tied to details of self-assembly or neural network models but are instead intrinsic to associative pattern recognition carried out through short-ranged interactions. 
\end{abstract}

\maketitle


Algorithms to recognize patterns in high-dimensional signals have made remarkable breakthroughs in the last decade, identifying objects in complex images and voices in noisy audio  \cite{Graves2013-qs,Krizhevsky2012-jc}. These algorithms are fundamentally different from earlier algorithms in that they often simulate a strongly interacting many-body dynamical system (i.e. a neural network) and exploit its emergent computational ability. Such emergent pattern recognition ability \cite{Hopfield1982-fb} raises the question of whether the dynamics of natural or engineered physical systems can directly show similar behavior. Pattern recognition would be of use to living organisms, which must often scan environmental signals for patterns that might be fuzzy, ill-defined, and can only be learned through examples rather than through a set of definitions \cite{Purvis2013-io,Levine2013-yd,Brubaker2015-ft}. 
Similarly, fuzzy recognition built into the internal physical dynamics of synthetic materials like sensors would allow to reduce the reliance on fragile electronic or neural systems for intelligent signal processing. Such an engineered `smart' material can exploit its dynamics to respond differently to, say, the chemical environment of Chicago top soil than to Peoria top soil, even if the individual chemical environments are themselves complex and highly variable and the differences are hard to define. 



Here, we show that the self-assembly dynamics of $N$ interacting molecular species can recognize patterns in those $N$ concentration levels. In the right window of physical parameters like temperature, binding energy and chemical potentials, even corrupted or incomplete patterns in the concentrations of $N$ molecular species are classified into one of several predefined classes through the assembly of designated structures. We use a model recently investigated in \cite{Murugan2015-ps}, where multiple self-assembling behaviors were programmed into a one set of self-assembling DNA molecules. Intuitively speaking, the pattern recognition described here exploits the physics of selective nucleation. In particular, we show that nucleation is sensitive to how closely correlations in concentration space reflect the physical proximity of those same species in self-assembled structures. 

The robust pattern recognition we find here is reminiscent of ``associative memory'' \cite{Hopfield1982-fb,Amit:1985tj,Hertz:1991ud} in neural networks. When programmed with examples of idealized patterns, such neural networks can then ``associate'' a new presented pattern with one of the idealized patterns, even if the presented pattern were corrupted or incomplete. Such networks have been extensively studied in statistical physics, theoretical neuroscience and computer science \cite{Amit1985-ls,MacKay2003-nq,
Hertz:1991ud}. They has been used for tasks such as handwriting recognition, and has served as a conceptual starting point for more complex models of pattern recognition.  

We show here that pattern recognition in self-assembly bears a strong similarity to continuous attractors models of neural networks \cite{Burak2012-bu,Chaudhuri2016-lh,Seung1997-ly,Wu2008-iw,Monasson2013-pn,Monasson2014-nu,Battaglia:1998bm,Seung2000-bk,Seung1997-ly}; these networks code for a continuum of states in short-ranged interactions and thus differ from the the original models of associative memory that code point attractors in long-ranged interactions. In fact, we show that self-assembly is mathematically closely related to associative memory in ``place cell'' network models \cite{Hopfield:2010tk,Battaglia:1998bm,Monasson2014-nu,Wu2008-iw} of the hippocampus that store representations, or maps, of multiple spatial environments through which an animal may want to move. We show that the storage of memories and their retrieval via pattern recognition in self-assembly are related to analogous processes in the place cell network. 

A particularly intriguing aspect of the connection between self-assembly and neural networks is that it relies on an extended quasi-particle approximation for the neural network. For $N$ binary neurons, this approximation reduces the configuration space from $2^N$ to a low dimensional attractor manifold of size $N$, corresponding to the quasi-particle center of mass. Such a reduction to a collective coordinate and its role in complex computational abilities has generated much interest \cite{Hopfield:2010tk,Battaglia:1998bm,Monasson2014-nu,Wu2008-iw,Hopfield2015-wt,Chaudhuri2016-lh,Burak2012-bu}. 

Thus our mapping between self-assembly dynamics and neural networks dynamics helps isolate exactly which aspects of each model are essential to pattern recognition ability.  We hope that such an abstracted understanding of requirements will stimulate further work on finding physical systems \cite{Fink:2001ki,Murugan2015-ps,Barish:2009te,wolynes088,wolynes095,wolynes114,wolynes116,wolynes381} whose dynamics can show the kind of learning behavior that has been successfully demonstrated over the last decade in machine learning research.

\begin{figure}
\centering
\includegraphics[width=\linewidth]{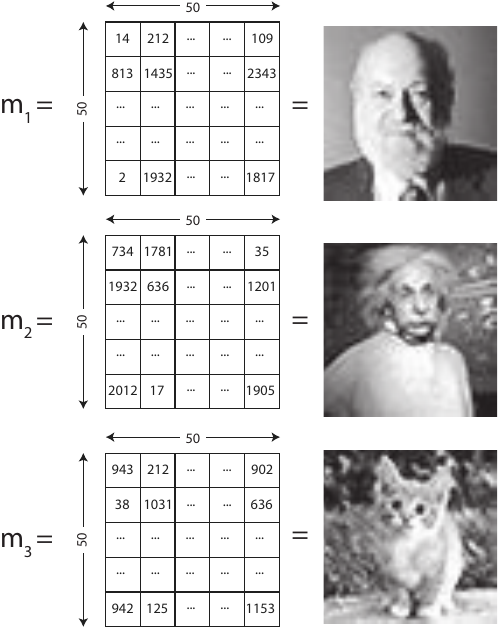}
\caption{A stored memory consists of a $2$-dim spatial arrangement of $N$ molecular species. Here, we show three distinct memories made of the same $N=2500$ species arranged as $50\times 50$ grids. Each species occurs exactly once in each memory. We will choose to represent these memories by gray scale images by associating each species with a unique gray scale value. Hence different memories correspond to different gray scale images in which each gray scale value is represented exactly once. Thus the three images here are permutations of the same $N=2500$ pixels. (Note: Many computer systems can only display $256$ distinct grayscale values.) }
\label{fig:memories}
\end{figure}

\begin{figure}
\centering
\includegraphics[width=\linewidth]{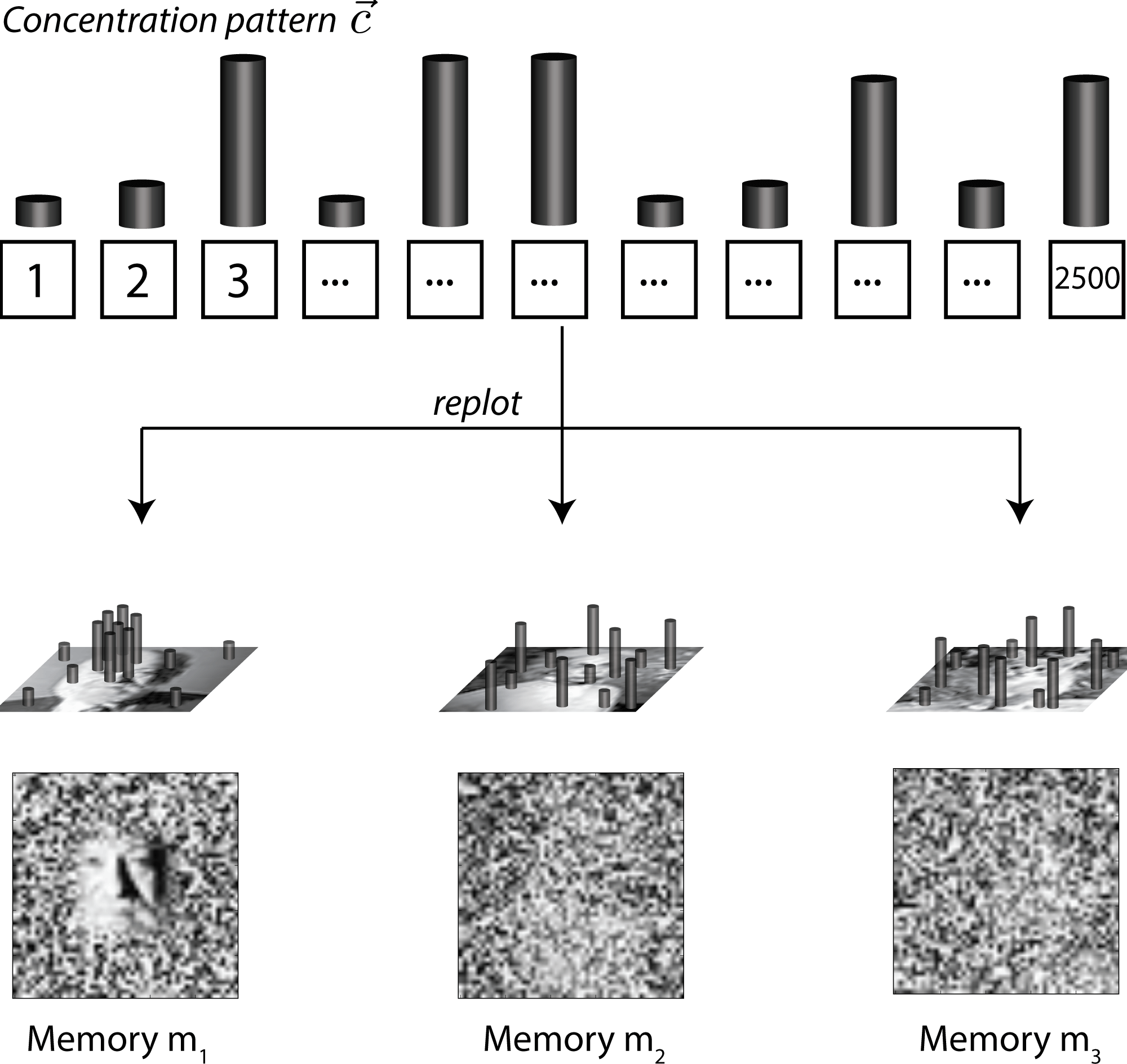}
\caption{A concentration pattern $\vec{c}$ has high overlap $\chi$ with memory $m$ if the species with high concentration in $\vec{c}$ are contiguous in the arrangement $m$. The pattern $\vec{c}$ shown has significant overlap $\chi \approx 1$ with the Leo memory but low overlap with the Einstein and cat memories. Here we visualize the overlap $\chi$ for each memory using a blurring mask described in the appendix. Pixels are assigned their true gray scale value if the average concentration of species in their neighborhood of area $a^*=100$ is high and a random value if the average concentration is low. Thus a clear region, like Leo's face, represents high overlap $\chi \approx 1$ and high average concentration. The high concentration species are scattered throughout the Einstein or cat memories, resulting in low average concentrations, low overlap and thus a blurry image.}
\label{fig:BlurryConcentrations}
\end{figure}

\begin{figure*}
\centering
\includegraphics[width=\linewidth]{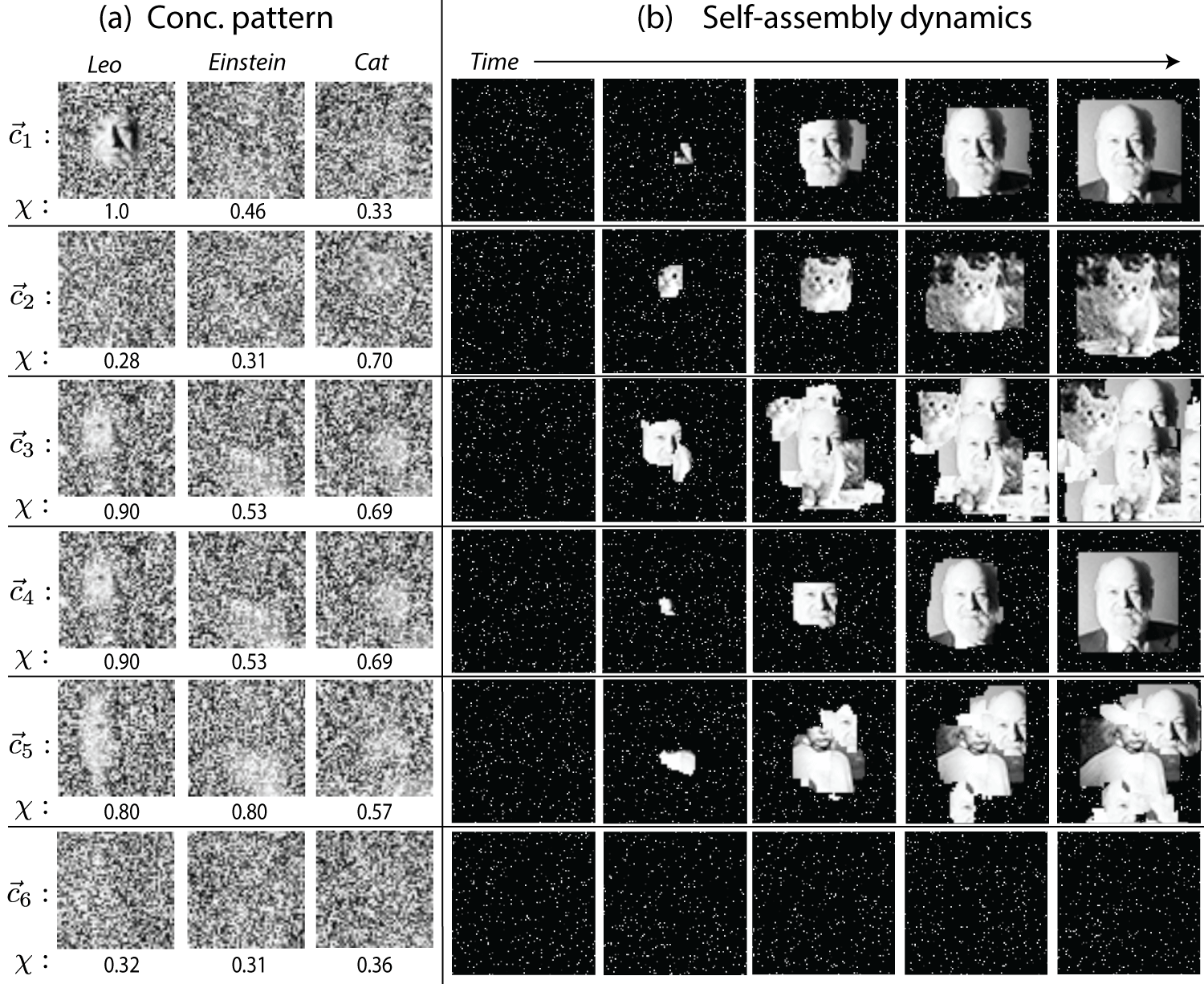}
\caption{Classification of patterns through self-assembly. Each row in (a) represents one concentration pattern shown in three representations as in Fig.\ref{fig:BlurryConcentrations}. The corresponding row in (b) shows the resulting self-assembly outcome. Concentration patterns $\vec{c}_1,\vec{c}_2$ with overlap with Leo or the cat respectively more than other memories assemble only that memory almost exclusively ($E_0 = 3 k T, \mu_{high} = 1.7, \mu_{low} = 1.9$). Patterns $\vec{c}_3$ and $\vec{c}_4$ have overlap with Leo and the cat are not as significantly different. When assembly is carried out with $\mu_{high} = 1.5$ as in $\vec{c}_3$, a chimeric mosaic of Leo and the cat is assembled, resulting in a failure to classify (row 3). However, in $\vec{c}_4$, the high concentration is reduced to $\mu_{high} = 1.7$; nucleation slows down for both the Leo and the cat but more so for the cat and only the Leo memory is evoked. Pattern $\vec{c}_5$ has an overlap of $\chi = 0.8$ with both Einstein and Leo; we find a mosaic of Leo and Einstein and classification fails. Finally pattern $\vec{c}_6$ has no significant overlap with any memory, giving no nucleation on the time scale of our simulations. (Note: One typical self-assembly outcome from Gillespie simulations combined digitally with image of monomers. For time stamps of the assemblies shown, see Appendix.)}
\label{fig:ClassificationSA}
\end{figure*}

\section{Pattern recognition in self-assembly}
\label{section1}
We assume we have $N$ distinct species of macromolecules. We begin with a few definitions:

\textbf{$\bullet$ A memory} is defined as a 2-dimensional arrangement of these $N$ species as a grid. Fig.\ref{fig:memories} shows three examples of such memories - referred to the `Leo',`Einstein' and `cat' memories, made of $N=50\times 50 = 2500$ species of molecules. As we will see below, memories play two roles in pattern recognition; (1) memories represent the physical 2-dimensional structure that is self-assembled in response to a concentration pattern. Thus memories $m_\alpha$ represent the discrete classes that environmental concentration patterns are classified into. A good pattern recognizer should produce only one self-assembled memory in response to an external concentration pattern; (2) the arrangement of the $N$ species in each memory $m_\alpha$ specifies the set of concentration patterns that must be classified as being of type $m_\alpha$.

\textbf{$\bullet$ A concentration pattern} is defined as a concentration vector $\vec{c}$ in which some number $s$ of species have a high concentration $c_{high}$ and the other $N-s$ species are at $c_{low}$. See Fig.\ref{fig:BlurryConcentrations}. 
For simplicity, we will restrict to the stored patterns to be made of binary concentrations $c_{high}$ or $c_{low}$ while describing our work.

$\bullet$ \textbf{Overlap between pattern and memory} The overlap $\chi_{a \alpha}$ is a measure of whether the species with high concentrations in pattern $\vec{c}_a$ are physically proximate (or contiguous) in the memory $m_\alpha$. We define $\chi_{a,\alpha}(x)$ to be the fraction of species in a square region of area $q$ centered at a position $x$ in memory $m_\alpha$ that have high concentrations in $\vec{c}_a$. Thus, if $\chi_{a,\alpha}(x) = 1$, all species in a region around $x$ in $m_\alpha$ have high concentrations in pattern $\vec{c}_a$. We define the overlap $\chi_{a,\alpha}$,
\begin{equation}
\chi_{a,\alpha} \equiv \mbox{max}_x \chi_{a,\alpha}(x) 
\end{equation}

To gain intuition, in Fig.\ref{fig:BlurryConcentrations}, we show the local overlap $\chi_{a,\alpha}(x)$ visually using a blurring mask, described in the Appendix; a blurry region indicates low overlap and hence lower concentrations than the average while an in-focus region indicates a region of high concentration.  For example, the species with high concentrations in the pattern in Fig.\ref{fig:BlurryConcentrations}a are all localized to a part of Leo's face in the 2-dim arrangement corresponding to Leo's memory. For $x$ in that part of Leo's face, $\chi_{a,\alpha}(x) = 1$ and thus $\chi_{a,\alpha} =1$; we say the pattern $\vec{c}_a$  matches Leo's memory $m_\alpha$ perfectly. On the other hand, these concentrations are widely distributed over the cat memory, as seen from the bar charts in Fig.\ref{fig:BlurryConcentrations}. Consequently, the cat image is entirely blurry.

Note that the overlap defined here serves as a generalization of the usual dot product to the case of continuous attractors \cite{Monasson2014-nu,Hertz:1991ud}; it is computed  by restricting to \textit{local} regions of size $q$ of the memory. See the Appendix for a discussion of the appropriate choice of area $q$ in computing the overlap; we use $q = 8 \times 8$ in Fig.\ref{fig:BlurryConcentrations}. 
 
$\bullet$ A concentration pattern $\vec{c}_a$ is said to \textbf{evoke} a memory $m_\alpha$ if, when species are supplied at concentration $\vec{c}_a$, a structure with molecules arranged as in memory $m_\alpha$ is self-assembled.

\subsubsection*{Storing memories}
To create self-assembly dynamics that can recognize concentration patterns, we first `program' the given set of idealized memories, such as those in Fig.\ref{fig:memories}, into the interactions of $N$ species. Inspired by associative memory found in neural networks \cite{Hopfield1982-fb, Hertz:1991ud}, we will take the binding affinities between the $N$ molecular species to reflect spatial relationship in the memories; that is, we assume that two molecular species have high affinity for each other if they occur next to each other in one of the memories   
\begin{eqnarray}
J_{ij} =  \begin{cases}
    -E_0 ,& \text{if } i,j \mbox{ are neighbors in any memory}  \\
    0,              & \text{otherwise}
\end{cases}\label{eqn:J2dim}
\end{eqnarray} 
Note that a given species $i$ will generally have distinct neighbors in different memories; hence the above prescription creates promiscuous interactions. Such a model of promiscuous interactions was studied in \cite{Murugan2015-ps} as a model of self-assembly with multiple target structures. \added{While analogous to self-assembly polymorphs, we are working in the limit of many distinct species. Consequently \cite{Murugan2015-ps} found that the number of different polymorphs that can be programmed in has a sharp limit, analogous to the sharp limit on the number of patterns that can be stored in large associative neural networks \cite{Hopfield1982-fb}.} Here, we will be interested in how such a promiscuous soup of molecules responds to various concentration patterns. 

\subsubsection*{Recognition through nucleation}

To understand how the promiscuous soup defined in Eqn.\ref{eqn:J2dim} can classify concentration patterns, we need to understand the selective nucleation of different memories by a concentration pattern. We find that a pattern $\vec{c}_a$ is much more likely to nucleate a memory $m_\alpha$ with high overlap $\chi_{a \alpha}$ than a memory $m_\beta$ with low overlap $\chi_{a \beta} <\chi_{a \alpha}$. 

To see this quantitatively, it is convenient to work with chemical potentials $\mu = - \log c$ of species instead of concentrations $c$. The average chemical potential of all species in a square region of area $q$ at position $x$ for memory $m_\alpha$ due to pattern $\vec{c}_a$ is given by,
\begin{equation}
\langle\mu\rangle(x) = \mu_{low} + \chi_{a\alpha}(x) (\mu_{high} - \mu_{low}) 
\end{equation}
Since $\chi_{a\alpha} = \mbox{max }_x \chi_{a\alpha}(x)$ and $\Delta \mu \equiv \mu_{high} - \mu_{low}<0$, the region of the memory with lowest average chemical potential (and thus greatest propensity to nucleate) has chemical potential $\langle\mu\rangle_m = \mu_{low} + \chi_{a\alpha}(\mu_{high} - \mu_{low})$. The rate $\Gamma$ of nucleating a seed in such a region can be calculated from classical nucleation theory by computing the energy barrier to nucleation. We find \cite{Murugan2015-ps},
\begin{eqnarray}
\Gamma(\chi_{a \alpha}) & = & 
    \exp\left(- \frac{E_0}{2 - \langle\mu\rangle_{m}/E_0}\right) \nonumber  \\
    &=&    \exp\left(- \frac{E_0}{f +   (\delta\mu) \; \chi_{a \alpha}   }\right)   
 \label{eqn:nucleationrate}
\end{eqnarray}
where $\delta\mu \equiv (\mu_{low} - \mu_{high})/E_0$ is a measure of the dynamic range of concentrations and $f \equiv 2 - \mu_{low}/E_0$ sets the growth rate of nucleated structures. Consequently, a pattern $\vec{c}_a$ will result in assembly of memories $m_\beta$ and $m_\alpha$ in the ratio $\frac{\Gamma(\chi_{a \beta} )}{\Gamma(\chi_{a \alpha} ) }$. Assuming $\vec{c}_a$ was to be classified as $m_\alpha$ because $\chi_{a \alpha} > \chi_{a \iota}$ for all stored memories $\iota$, we define the error rate in classification as, 
\begin{equation}
\eta =  \frac{\sum_{\iota \neq \alpha} \Gamma(\chi_{a \iota} )}{\Gamma(\chi_{a \alpha} ) } \label{eqn:error}
\end{equation}
where the sum is over all stored memories except $\alpha$.

With this definition of error rate $\eta$, we see that classification fundamentally relies on the nucleation rate $\Gamma$ showing large variation as a function of the overlap $\chi$. 

\subsection*{Response to patterns}

What is the response of a soup of promiscuously interacting particles (defined by Eqn. \ref{eqn:J2dim}) to different concentration patterns $\vec{c}$? How frequently are patterns incorrectly classified? How do the error rate $\eta$ and time to classify $\Gamma$ depend on parameters of the theory? 

To answer these questions, we simulated self-assembly for a set of illustrative cases of concentration patterns shown in Fig. \ref{fig:ClassificationSA}. In each case, we took a set of $N$ species with interactions as defined by Eqn. \ref{eqn:J2dim} and set the concentrations to be a chosen pattern $\vec{c}$. We assume for simplicity that concentrations are held fixed during self-assembly, as in a grand canonical ensemble. We simulate self-assembly using the Gillespie algorithm detailed in the Appendix and report on the mix of assembled structures produced.


$\bullet$  Pattern $\vec{c}_1$ in Fig. \ref{fig:ClassificationSA} was taken to perfectly match a part of the Leo memory ($\chi = 1.0$). That is, the concentrations of $s=364$ species that form a clump around Leo's left eye were taken to be $c_{high}$ and the rest $c_{low}$. The high concentration species were randomly spread out and not contiguous in the other memories. We observed rapid nucleation and subsequent assembly of the Leo memory exclusively. 

$\bullet$ Pattern $\vec{c}_2$ in Fig. \ref{fig:ClassificationSA} imperfectly matches the cat memory ($\chi = 0.7$) and poorly matches the others. Such an imperfect pattern is nevertheless correctly classified by self-assembly since the Leo and Einstein overlaps are much lower ($\chi = 0.28 ,0.30$). However, the nucleation rate $\Gamma$ is lower than for $\vec{c}_1$ since $\chi= 0.7$ has significantly lower average concentration in the nucleation region relative to $\chi = 1.0$. That is, the time to classify has increased.

$\bullet$  Patterns $\vec{c}_3$ and $\vec{c}_4$, which are structurally identical, have higher overlap $\chi=0.9$ with Leo than with the cat $\chi = 0.69$. But if we speed up nucleation by setting $\mu_{high} = 1.5$ for $\vec{c}_3$, we find a high error rate. Many cat memories are nucleated at sites on the boundary of growing Leo memories (`heterogeneous nucleation').  These errors can be eliminated almost entirely by increasing $\mu_{high} = 1.7$ for $\vec{c}_4$. Increasing $\mu_{high}$ lowers nucleation rates for both Leo and the cat but the relative rate increases. Thus we trade longer classification time for lower error rate. This speed-accuracy tradeoff is discussed in more detail below.

$\bullet$  Pattern $\vec{c}_5$ has near identical overlap $\chi=0.8$ with both the Leo and Einstein memories. Such an ambiguous pattern can exist only because the Einstein and Leo memories themselves were designed resemble each other in a region near Einstein's shirt and Leo's face. Pattern $\vec{c}_5$ has high concentrations in such a region, and consequently, pattern $\vec{c}_5$ evokes a mixture of Leo and Einstein.

$\bullet$ Finally, pattern $\vec{c}_6$ involved $s$ high concentration species that were not significantly grouped together in any memory. Such a pattern has low overlap with all memories and does not evoke (i.e. nucleate) any memory within the duration of the simulation.




\subsubsection*{Error rates can be lowered through slower nucleation}

As mentioned when discussing the response to patterns $\vec{c}_3$ and $\vec{c}_4$, nucleation dynamics can be accelerated by lowering $\mu_{high}$. However, lowering $\mu_{high}$ beyond a point becomes counter-productive in terms of error rate; e.g., if $\mu_{high} \approx \mu_{low}$, all patterns $\vec{c}$ are indistinguishable from uniform concentrations. If time to nucleate the structure were no concern, what is the optimal value of $\mu_{high}$ that would give the  smallest error rate? From Eqn. \ref{eqn:nucleationrate} and \ref{eqn:error}, we find that to the minimize the error rate in distinguishing two patterns with overlaps $\chi_r$ (with the `right' memory) and $\chi_w$ (with the `wrong' memory), we should choose
\begin{equation}
\mu_{low} = \mu_{high}\left(1 + \frac{1}{\sqrt{\chi_r \chi_w}}\right)
\end{equation}

\subsubsection*{Trade-off: Time vs error vs overlap difference}

We also saw that while $\vec{c}_3$ and $\vec{c}_4$ were the same pattern structurally, the rapid dynamics in response to $\vec{c}_3$ resulted in a proliferation of errors. This illustrates a fundamental trade-off between error rate and classification time. We can attempt to speed up classification (i.e., increase the nucleation rate $\Gamma_{r}$ of the `right' memory with highest overlap) by increasing the high concentration (i.e., lowering $\mu_{high}$) or lowering the binding energy $E_0$ (or equivalently raising the temperature). However, this causes the nucleation rate to increase for all memories with any overlap $\chi$ and, in fact, the error rate $\eta$ then increases.  Thus, we can trade off lower classification time $1/\Gamma_r$ for higher classification error $\eta$ and vice-versa. 

\begin{figure}
\centering
\includegraphics[width=0.9\linewidth]{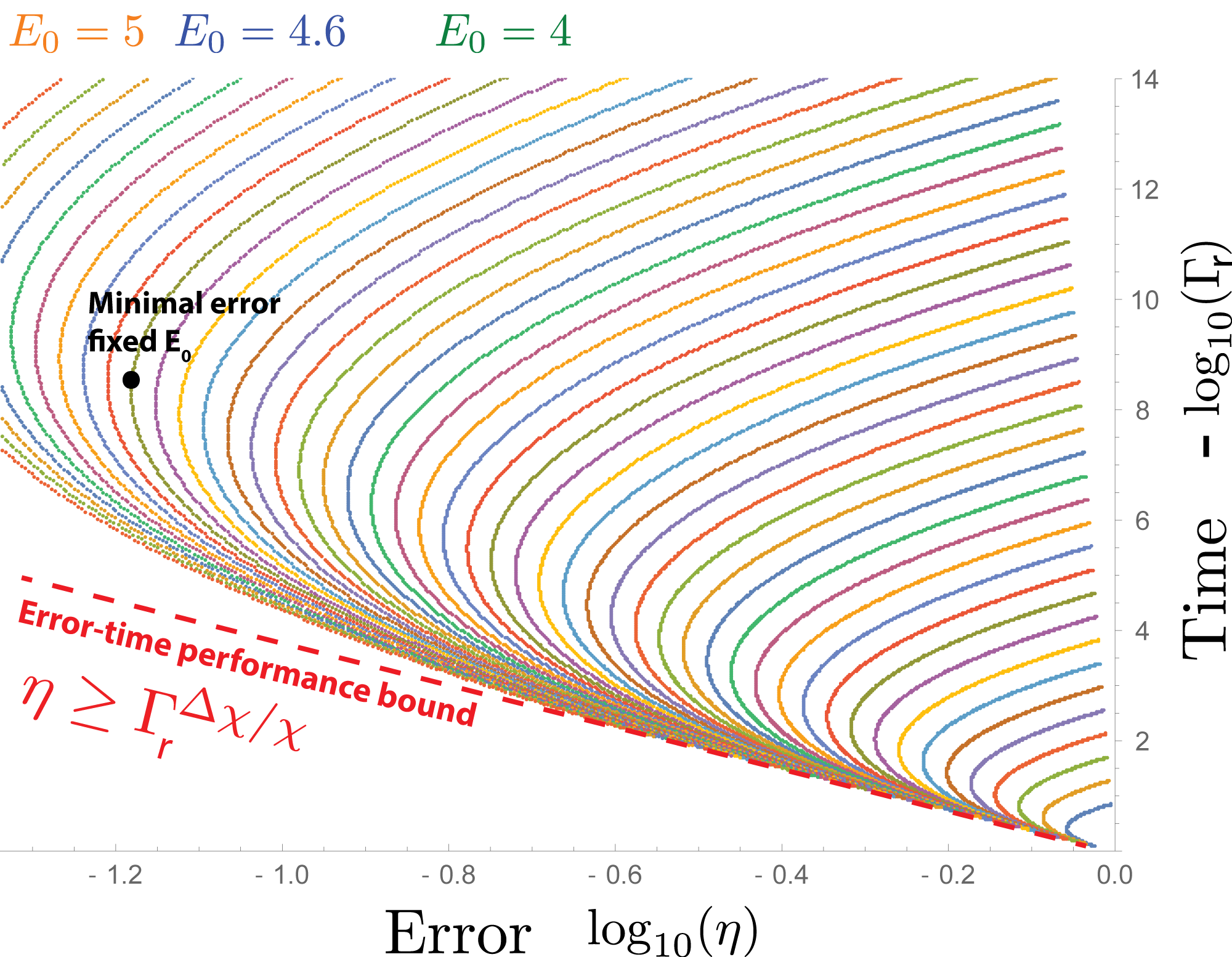}
\caption{Trade-off between classification time and classification error \added{based on nucleation rates (Eqns.\ref{eqn:nucleationrate}, \ref{eqn:error}).} We evaluated classification time $\Gamma_r$ and error $\eta$ for pattern $\vec{c}_3$ of Fig.\ref{fig:ClassificationSA} with overlap $\chi = 0.9$ with Leo and $\chi = 0.69$ with the cat for many values of binding energy $E_0$ and chemical potential $\mu_{high}$. We find an absolute lower bound on error for any fixed classification time (red line) given by $\eta = \Gamma_r^{\Delta \chi/\chi_{\mbox{cat}}}$ (See Eqn.\ref{eqn:tradeoff} for a derivation). \added{If $E_0$ is constrained to be a fixed value as would be common experimentally \cite{Ke2012-lm,Wei2012-ox}, the trade-off curves parameterized by $\mu_{high}$ are shown by lines of different colors. The black dot shows the minimal error for a fixed $E_0$ if time were immaterial.}}
\label{fig:tradeoff}
\end{figure}

To understand this trade-off, in Fig \ref{fig:tradeoff} we made a scatter plot of classification time $1/\Gamma_r$ versus the error rate $\eta$ as the parameters, $E_0$ and $\mu$, are varied. We find a fundamental error-time performance bound; given a fixed time, the error rate is bounded from below. In fact, we can derive an analytic formula for this bound from Eqns.\ref{eqn:nucleationrate} and \ref{eqn:error} by extremizing over parameters; we find that
\begin{equation}
\scalebox{1.2}{$\eta \geq \Gamma_r^{\frac{\Delta\chi}{\chi_{w}}}$} \label{eqn:tradeoff}
\end{equation}
where $\Delta\chi = \chi_r - \chi_w$ and $\chi_r$ is the largest overlap (i.e., overlap with the `right' memory) and $\chi_w$ is the next highest overlap (i.e., overlap with the dominant `wrong' memory). 

Thus, the minimum achievable error rate always falls off if the classification time is increased (i.e., nucleation rate $\Gamma_r$ is decreased). But the rate at which it drops off is much slower for a pattern that has similar overlap with different memories (i.e., with small $\Delta\chi$). Thus patterns with closer overlap necessarily take longer to distinguish.

\subsubsection*{Pattern sparsity sets dynamic range of concentrations}
The requirement that a random pattern with $s$ high concentration species not evoke a random memory imposes a constraint on the dynamic range of concentrations $c_{high}$ and $c_{low}$. 

To see this, note that random patterns and memories will generically have overlap $\chi \sim s/N$. Due to the underlying discrete nature of our structures, the nucleation rate $\Gamma$ effectively reaches its highest value  when $f + (\delta \mu) \chi \sim  0.5 $ (or equivalently $\langle \mu \rangle_{m} \sim 1.5 E_0 $); at this point, the critical nucleation seed has area $\sim 2 \times 2$ and thus the nucleation barrier effectively disappears. Combining this inequality with the requirement that the seed grows, $f_{growth}> 0$, we find that $$\mu_{high}/E_0 > 2 - N/2s.$$ Thus, the sparsity $1-\frac{s}{N}$ of the pattern $\vec{c}_a$ sets an upper bound on $c_{high}$.

\subsubsection*{Capacity: Maximal number of patterns}
Finally, we have only considered three patterns in our simple example and found that the self-assembly properly forms the intended structure in the right regimes of $\chi,\mu,E_0$.  

However, as shown in \cite{Murugan2015-ps}, once the number of programmed memories exceeds a threshold, $m_c$, the capacity of the system, promiscuous interactions lead to a combinatorial explosion of chimeric structures. The capacity $m_c$ was found to scale with the size of the memories as $N^{1 - 2/z}$ where $z$ is the coordination number of the structure. Even a pattern with perfect overlap $\chi = 1$ with, say Leo, cannot faithfully evoke the Leo memory near or past this capacity transition; chimeric combinations of different stored memories will be assembled in response.

\subsubsection*{Inverse design: memories from patterns}
In the above discussion, we started with the notion of memories -- a $2d$ spatial arrangement of $N$ species -- which defined a whole set of concentration patterns that match it. Instead, if one is given a set of concentration patterns $\vec{c}_a$, how can one design memories and thus interactions $J_{ij}$ such that each pattern evokes a unique memory structure?  

We can define a memory $m_\alpha$ for each pattern by taking $s$ of the species that are at high concentrations and design a structure that puts those $s$ species together as a contiguous region in a random permutation in one part of the structure. Species with low concentrations could be placed in a random arrangement surrounding this seed. With such a prescription, each given concentration pattern $\vec{c}_a$ would reliably evoke memory $m_i$ since the odds of the same seed occurring in another memory $m_\beta$ would be negligible below the capacity threshold. 

However, note that memory $m_\alpha$ will necessarily be evoked by patterns unrelated to $\vec{c}_a$ since patterns localized in other regions of $m_\alpha$ will also evoke it. These other unavoidable patterns associated with $m_\alpha$ are reminiscent of spurious patterns in neural network models; here, these spurious patterns are a inevitable consequence of using short ranged interactions to program continuous attractors instead of point attractors. We discuss this distinction further in the conclusions.

\section{Relationship between self-assembly and place cell neural networks}

\added{We now show that the self-assembly model is able to recognize concentration patterns because of a remarkable similarity to place cell neural networks that store spatial maps of environments.
Relying on a simple model of place cells based on recent experiments, we show the connection at two levels; (1) At the level of structure of the interactions, both models create an internal representation of external spatial relationships. (2) At the level of dynamics, pattern recognition in self-assembly is related to a collective coordinate approximation of place cell dynamics.}

\begin{figure}
\centering
\includegraphics[width=0.9\linewidth]{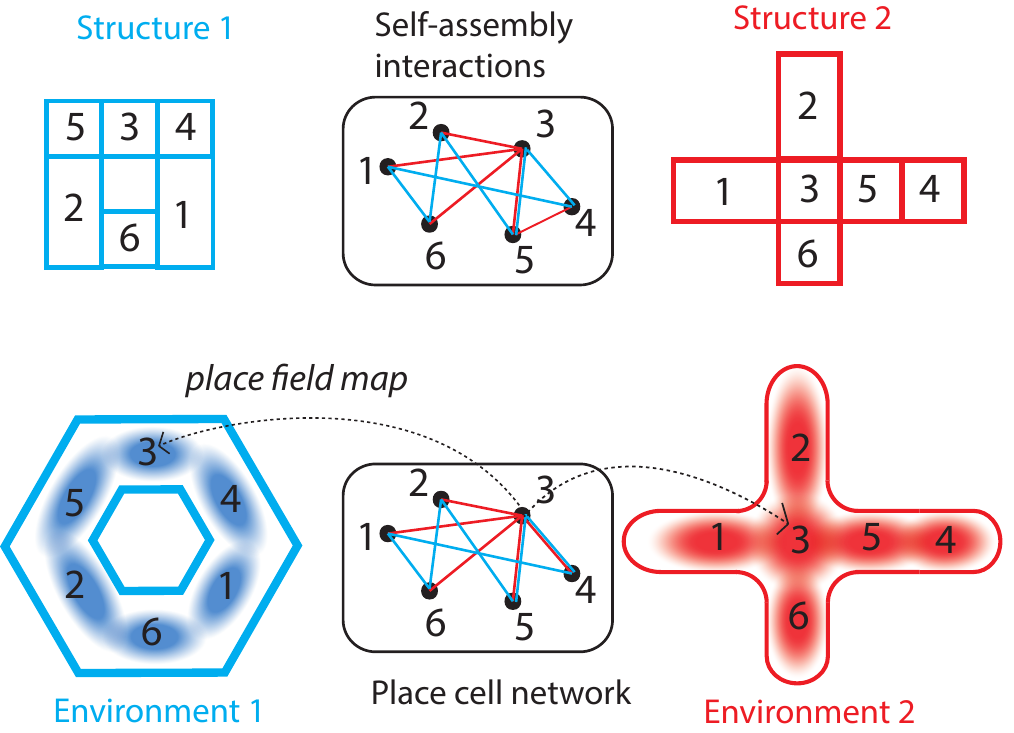}
\caption{Programming self-assembling particles with multiple target structures is analogous to programming a place cell neural network with multiple spatial environments. Both systems are programmed with internal representations (by tuning particle interactions or neuronal connections) of external spatial relationships (between particles or place fields). The self-assembling particles can then coherently grow any select stored structure from a seed; the neural network can `mentally explore' \cite{Hopfield:2010tk} an environment selected by initial conditions by traversing through it.}
\label{fig:Hopfield2Rooms}
\end{figure}

\subsection{Structure: Internal representation of external spatial relationships}

As shown in Fig \ref{fig:Hopfield2Rooms}, in self-assembly, the affinities between different species code for the spatial proximity of species in different target structures. In particular, for each of the $\alpha = 1, \ldots , m$ target structures, we assume that the interaction matrix between species is,
\begin{eqnarray}
 J^\alpha_{ij} = \begin{cases}
    1 ,& \text{if }  |f_\alpha(i)-f_\alpha(j)|<d  \\
    0,              & \text{otherwise}
\end{cases}
\label{eqn:JSA}
\end{eqnarray}
where $f_\alpha(i)$ is the spatial location of species $i$ in structure $\alpha$ and $d$ is the interaction range. We take the total interaction matrix to be,
\[
J^{tot}_{ab} = \sum_\alpha J_{ab}^\alpha
\label{eqn:JSAtot}
\]

Recent neuroscience experiments reveal a similar picture \cite{Colgin2010-rd, Jezek2011-sz, Wills2005-hz} for how mammals represent memories of spatial environments in the hippocampus. When a rodent has been exposed to a spatial environment (e.g., a room), place cells in the hippocampus are each assigned a `place field', i.e., a small spatial region of that environment. A place cell fires only when the rodent is in that cell's `place field' region of the environment.  As the rodent moves through the environment, place cells with nearby place fields fire concurrently when the rodent is in the overlap region of two place fields. Thus place cells with nearby place fields might strengthen their synaptic connections in a Hebbian fashion (cells that 'fire together, wire together') while connections between cells with distant place fields are not facilitated. (`Place cells' may arise as effective modes of 'grid cells' in the entorhinal cortex but we will use the effective place cell description here.) 
 
Thus after being exposed to one environment $\alpha$, place cells in the hippocampus will have connectivity,
 \begin{eqnarray}
 J^\alpha_{ij} = \begin{cases}
    1 ,& \text{if }  |f_\alpha(i)-f_\alpha(j)|<d  \\
    0,              & \text{otherwise}
\end{cases}\label{eqn:JNN}
 \end{eqnarray}
where now $f_\alpha(i)$ is the location of the center of neuron $i$'s place field in environment $\alpha$ and $d$ sets the range of interactions between place cells. Thus, Eqn.\eqref{eqn:JNN} says that place cells whose place fields in environment $\alpha$ overlap are wired together.

Exposing the rodent to a different environment \cite{Kubie:1991vq}, say $\beta$, creates another seemingly random assignment of place fields for the same place cells that appear to bear no relationship to place field arrangements in environment $\alpha$ Exposure to environment $\beta$ will create a new set of neuronal connections between place cells, in addition to those created by environment $\alpha$. We assume \cite{Monasson2014-nu,Hopfield:2010tk,Battaglia:1998bm} that the connections between place cell neurons after being exposed to $m$ environments are simply a sum of each individual environment's contribution, i.e.
\begin{equation}
J^{tot}_{ij} = \sum_\alpha J_{ij}^\alpha
\label{eqn:JNNtot}
\end{equation}

In both constructions, Eqn.\ref{eqn:JSA} and Eqn.\ref{eqn:JNN}, the internal interaction matrix reflects external spatial relationships. In self-assembly, the affinity matrix $J^{tot}_{ij}$ between chemical species codes for the spatial relationships of those species in the different stored structures. In the hippocampus, the synaptic connectivity $J^{tot}_{ij}$ codes for the spatial relationships of place fields in the different stored environments. In fact, \cite{Curto:2008vr} showed that the geometry and topology of stored environments can be mathematically reconstructed from the pooled matrix $J^{tot}_{ij}$. 


\subsection{Dynamics : Inhibition and attractors}

\begin{figure*}
\centering
\includegraphics[width=\linewidth]{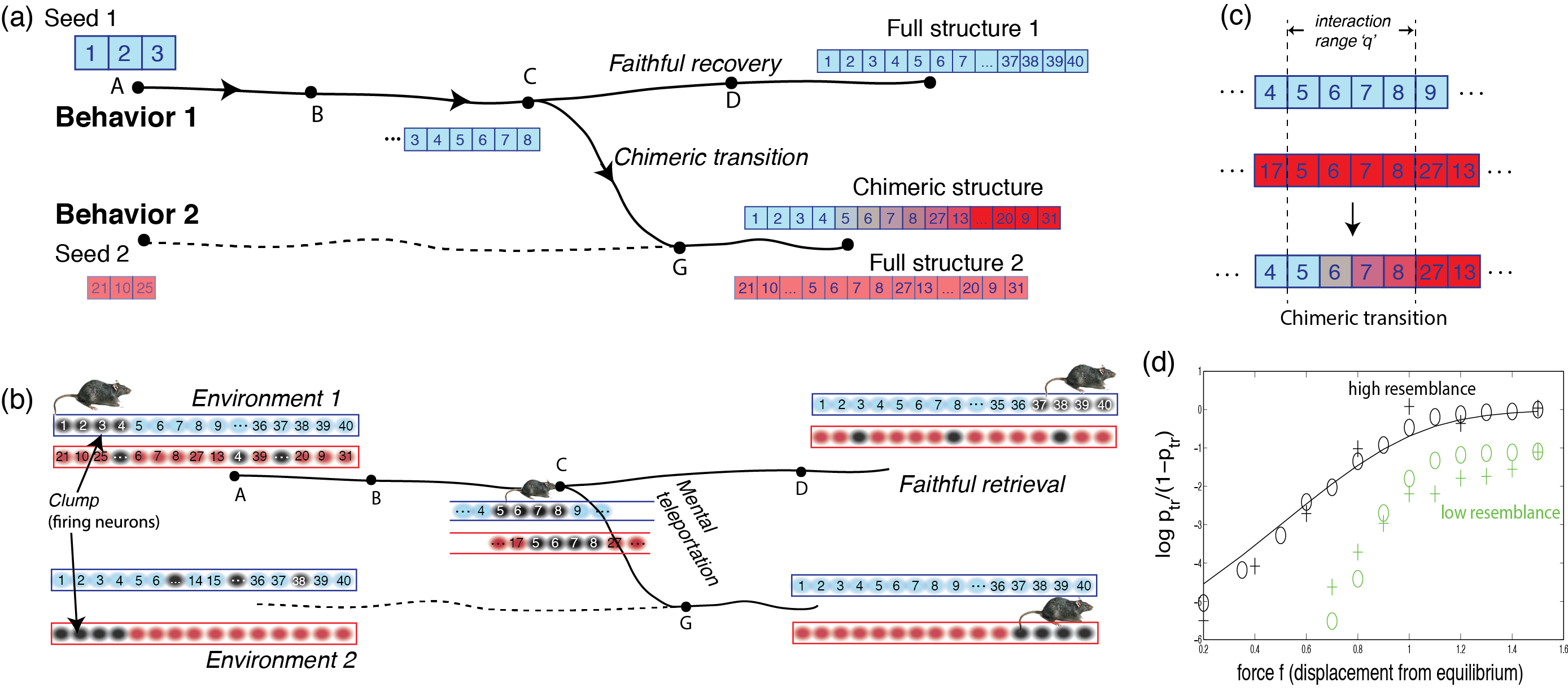}
\caption{Self-assembly and neural networks programmed with two memories (red and blue) with high resemblance in a region leads to similar failure to classify due to chimeric structures or mental teleportation. (a) A fragment of the programmed blue structure, growing through self-assembly can transition to growing as the red structure when growing through a region of high resemblance shown in (c). The result is a  half blue-half red chimeric structure. (b) If neural activity is initialized as a clump at one end of the blue environment, the rat should be able to mentally explore the blue environment by driving the clump from  left to right in environment 1 (e.g., path planning \cite{Pfeiffer2013-qn, Ponulak2013-op}). However, the clump can transition to being a clump in environment 2 (red) when passing through a region of high resemblance (state C). (d) The probability of chimeric transitions in self-assembly and teleportation in neural networks grows with the force $f$ used for retrieval and with local resemblance between stored memories (circles - self-assembly, crosses - neural networks, solid - \added{barrier crossing formula Eqn.\ref{eqn:barriereqn}}) }
\label{fig:TrajectoryBranches}
\end{figure*}

\added{Despite the similarity of interaction matrices, the state space of these systems is in principle very different. The energy of a self-assembled structure depends on the spatial ordering of different species that occur in the structure; on the other hand, a state of the neural network is determined by which set of neurons is firing and there is no corresponding ordering information. How could the states and dynamics map on to each other?} 
 
\textbf{Collective coordinate of the neural network}
Long-range global inhibition found in the hippocampus limits the total number of neurons that can be active at once. Earlier theoretical work \cite{Monasson2014-nu,Hopfield:2010tk,Battaglia:1998bm, Wu2005-sw, Wu2008-iw} showed that the balance between short-ranged excitation and long-range  inhibition causes neural activity to condense into a contiguous `clump' of some fixed size $l$, i.e., $l$ neurons with place fields in one region of space fire while all others are off. Such a stable clump --- a collective coordinate for neural activity --- corresponds to a specific rodent position in the physical environment.  This constraint is analogous to working at fixed magnetization in a ferromagnetic Ising model; the firing neurons condense into a clump of activity to minimize its surface area.

\begin{figure}
\centering
\includegraphics[width=1\linewidth]{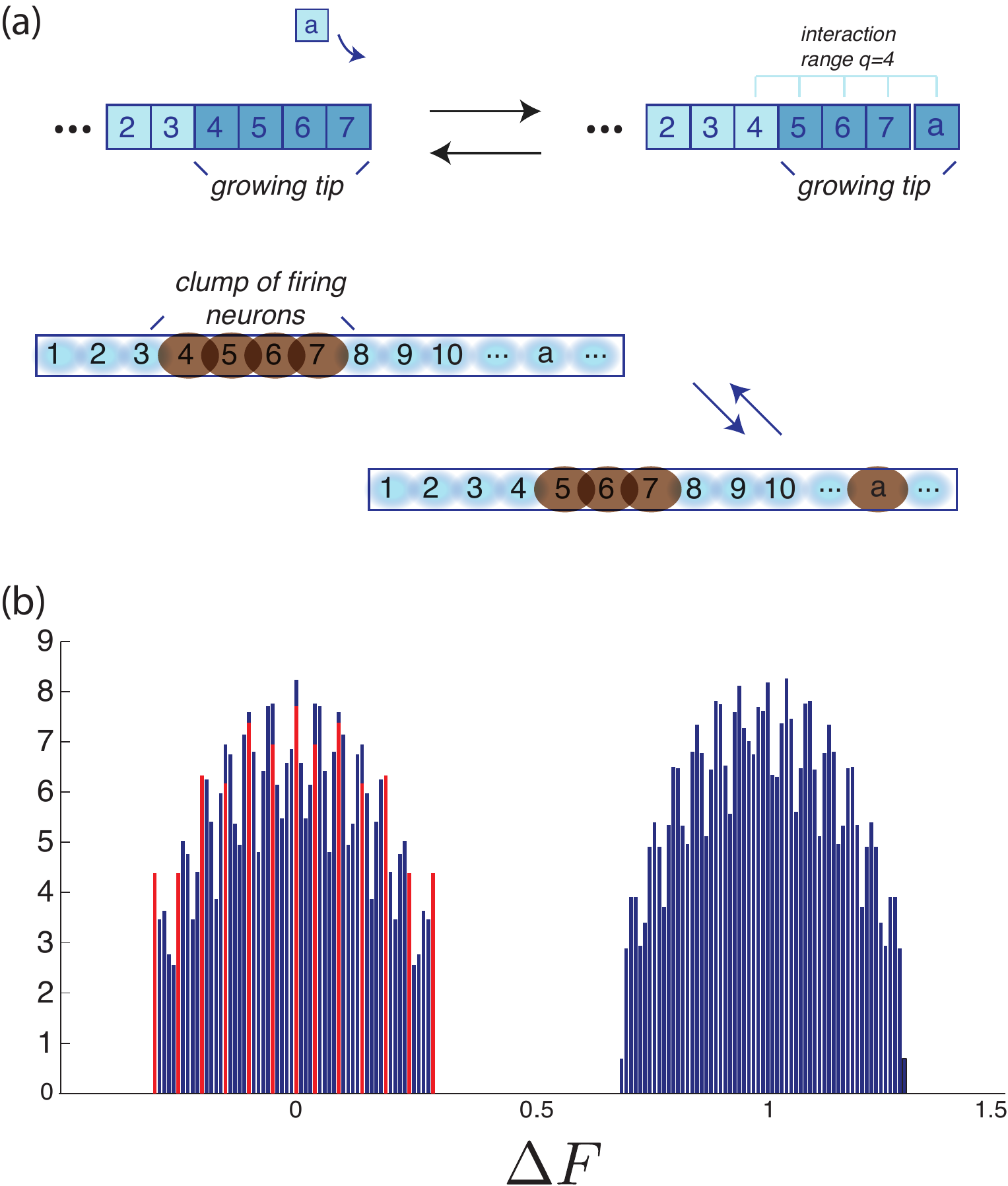}
\caption{(a) All transformations of the growing tip of a self-assembling structure and the low energy transformations of the clump of place cell activity can be put in a $1-1$ correspondence. The clump also has higher energy transformations which do not correspond to any transformation of the growing tip of structures. (b) We computed the histogram of free energy changes along trajectories of length $k=6$, starting at the same state, for both systems (place cells in blue, self-assembly in red). The low energy trajectories of the clump are matched by trajectories of self-assembly. (Driving force $f = 0.05$ in these simulations.) 
\label{fig:SI_Clump_Tip_correspondence}}
\end{figure}

The clump can be driven to move around the network, which is interpreted as mental exploration of a spatial environment. See Fig. \ref{fig:SI_Clump_Tip_correspondence}. The description of neural activity in terms of such a collective coordinate, e.g., the center of mass of the clump, simplifies the problem, reducing the configuration space from the $2^N$ possible states of $N$ neurons to $N$ possible center of mass states of the clump along the continuous attractor \cite{Wu2008-iw}.

\textbf{Simple model of two memories in 1-dimension}

While our focus in the previous section was nucleation, here we study the similarities in growth dynamics of self-assembly to clump motion in attractor neural networks. To gain intuition, we contrast the dynamics of self assembly and place cells when they both encode the same pair of $1$-dimensional structures or environments.
As shown in Fig.\ref{fig:TrajectoryBranches}, the first memory in both cases is set to $1-2-\ldots-N$ without loss of generality. The second memory is a random permutation but with a region of some size that is in common with memory 1. We assume that interactions extend over $q=4$ nearest neighbors (range $q$ sets parameter $d$ in Eqns.\ref{eqn:JSA} and \ref{eqn:JNN}).

That is, in both cases, the total interaction matrix is the sum of interaction matrices corresponding to two memories (i.e., two structures made of the same species or two spatial environments covered by the same place cells):  
\[ 
J^{tot}_{ab} = J^{(1)}_{ab} + J^{(2)}_{ab} \] 

The central question is whether dynamics based on $J^{tot}_{ab}$ can effectively show behaviors corresponding to only one of the pooled matrices, say $J_{ab}^{(1)}$, chosen by initial conditions. 

For self-assembly, we ask if we can selectively grow one of the stored structures (blue) from a seed belonging to that structure using a supply of free components as shown in Fig. \ref{fig:TrajectoryBranches}.
The two structures programmed in possess a region of overlap with strong similarity in a region of length $l=4$. At first, the self-assembly process will begin to grow the first structure. But if the common region is comparable to the range of interactions ($q=4$), growth may switch and continue growing as the red structure, thereby producing a blue-red chimera. We computed the probability of a chimera for different degrees of resemblance and as a function of the driving force completing the reaction. See circles in Fig. \ref{fig:TrajectoryBranches}.  (Details of the 1d self-assembly model are provided in the appendix.)

For place cell networks, we ask whether the rodent's neural activity clump can traverse (i.e, `mentally explore' \cite{Hopfield:2010tk}) one environment (say, blue in Fig.\ref{fig:TrajectoryBranches}b), from one end to the other, without being affected by the other. Or might a `mental teleportation' event in a region of high resemblance of place field arrangements (Fig.\ref{fig:TrajectoryBranches}c) spontaneously destroy the clump in one environment and reform the clump in the other environment?  Similar transitions have been seen experimentally in rats \cite{Jezek:2011ib} and in simulations without forces \cite{Monasson2013-pn,Monasson2014-nu}. Details of our simulations including a force are in the Appendix.

We find that the probability of a teleportation event, shown as crosses in Fig.~\ref{fig:TrajectoryBranches}d is determined by place field resemblance in the overlapping region as well as the driving force, in close agreement with the self-assembly results for the same pair of $1$-d permutations (circles in Fig.~\ref{fig:TrajectoryBranches}d). 

We are able to accurately capture such chimeric/teleportation behavior of both systems by modeling chimeric/teleportation transitions as a barrier crossing process under a driving force. For the case of near-perfect resemblance (black circles and crosses in Fig.~\ref{fig:TrajectoryBranches}(d)), we find 
\begin{equation}
\frac{p_{trans}}{1-p_{trans}} \sim \frac{e^{ f} + e^{ B}}{e^{f}+ 1}
\label{eqn:barriereqn}
\end{equation}
where $B$ is the energy barrier along the chimeric transition (i.e., segment $CG$ in Fig.~\ref{fig:TrajectoryBranches}), in good agreement with simulations of both neural networks and self-assembly. $B$ is the energy barrier along the chimeric transition (i.e., segment $CG$ in Fig.~\ref{fig:TrajectoryBranches}) due to imperfect resemblance; for the black curve, we used a resemblance with only one mismatch at the end of the common region in Fig.\ref{fig:TrajectoryBranches}(c) and thus $B = E_0$ (one missing bond). $B = 0$ if the resemblance is perfect over a region of size $q$. 


This example of two memories indicates that the clump of neural activity is analogous to the growing tip of the self-assembling structure. We explore this relationship quantitatively below. 






\textbf{Fluctuation spectrum of growing structures and the neural clump}

The place cell network model and self-assembly bear strong similarity because transformations of the growing tip of length $\sim q$ of a structure can be mapped to low energy transformations of an activity clump of size $l \sim  q$ with the same composition as the growing tip. We find that the mapping works to the extent that the collective coordinate approximation is appropriate. 

As shown in Fig.~\ref{fig:SI_Clump_Tip_correspondence}, all the transformations of self-assembly map onto transformations of the clump in a one-to-one manner, and the energies match up as well.  For example, the free energy change in adding a species $a$ to a growing tip is equal to the free energy change in turning on neuron $a$ and turning off the weakest neuron in the clump; both transformations have a free energy change $\Delta F =  -f +w E_0$ where $w$ is the number of zero energy bonds made by the new molecule (neuron) $a$ with the growing structure (moving clump) and $f$ is the driving force on the clump / on assembly (see Appendix).

However, the neural activity has many other transformations that make it discontiguous and thus do not correspond to addition of molecules in self-assembly. But, by definition, these fluctuations break the collective coordinate approximation and they are generally of higher energy. As a result, such energetically expensive transformations (which leave the attractor manifold) are likely to be followed by transformations which subsequently restore contiguity to the clump (and thus return the system to the attractor manifold). 

To quantify such corrections to the collective coordinate approximation, we numerically compute the free energy change associated with all possible sequences of $k$ molecule additions and $k$ neuron flips -- i.e., we found the free energy changes for all trajectories of length $k$ for both systems. For low $k \sim 1 ,2$, we find that the clump indeed has several transformations that do not correspond to any transformation of the tip and are only of slightly higher energy. However, as $k \sim O(q)$, we find that such extra transformations are pushed to significantly higher energies; the histogram of free energy changes for trajectories of length $k = 6$ are shown in Fig.~\ref{fig:SI_Clump_Tip_correspondence}(b) (blue - place cells, red - self-assembly).  We see that the low energy part of the spectrum matches up between the two systems.





\section{Similarities and differences}

\textbf{Multiplicity of species and place fields}

The self-assembled structures we considered here had only one species of each type (e.g., in Fig. \ref{fig:memories}, each pixel has a unique gray scale value). Such a heterogeneous limit of self-assembly \cite{Hedges2014-af,Murugan2015-wb,Jacobs2013-rk,Jacobs2015-jt, Haxton2013-bm, Whitelam2012-nt} describes  macromolecular complex assembly \cite{Levy:2006wv} and, in particular, synthetic DNA brick assemblies \cite{Ke2012-lm,Wei2012-ox} where each DNA strand is used exactly once in a structure. However, such an assumption is not essential to our model; as long as multiple copies of a species occur far enough apart in the memories, the robust pattern recognition seen here will continue to hold \cite{Murugan2015-ps}.

Similarly, recording experiments on the hippocampus show that that the same place cell might have multiple place fields in a given environment. Much like with self-assembly, the behavior of place cell network models remains unchanged provided the place fields of a given place cell are widely separated in a given environment \cite{Hopfield:2010tk,Monasson2013-pn}.

\textbf{Non-specific interactions}

In self-assembly, we assumed that we can code arbitrary spatial arrangements specified by memories $m_\alpha$ into the chemical affinities between $N$ species and that there are no other undesired non-specific interactions between the $N$ species.  While DNA sequence-based interactions \cite{Ke2012-lm,Wei2012-ox} are highly programmable, non-specific interactions can never be completely avoided. The effect of such interactions was shown to be small provided the strength of non-specific interactions was below a threshold logarithmic in the number of species $\sim \log N$  \cite{Hedges2014-af,Jacobs2013-rk,Jacobs2015-jt,Murugan2015-ps}. 

In place cell networks, sufficiently large heterogeneity in the  strengths of neural connections has been shown to destabilize continuous attractors and break it into a series of discrete attractors \cite{Seung1997-ly, Chaudhuri2016-lh}. 

\textbf{Classification vs memory storage}

We relied on nucleation in self-assembly to classify concentration patterns. The condensation of the activity bump in place cell networks into a single environment's attractor serves as an analogous classifier of inputs. However, continuous attractors in neural networks are used often as short-term memory storage devices in addition to being a classifier; the precise position of the activity clump within the attractor manifold is interpreted as storing a continuous number \cite{Chaudhuri2016-lh} that can later be retrieved or reset to another state. An analogous function does not exist in self-assembly.

\textbf{Self-assembly as the path integral of neural activity}

The relationship between neural networks and self-assembly is fundamentally a relationship between short trajectories in the two systems; moving the neural activity bump from point A to B corresponds to growing the structure between these two states. 

However, a major difference is that while the bump's final state at $B$ has no memory of its past (e.g., being at $A$), the self-assembled structure encodes the whole history of the bump's movement in its internal structure. For example, in Fig.\ref{fig:TrajectoryBranches}, the final state of the teleported clump in the red environment has no memory of undergoing the teleportation; however, the corresponding half-blue half-red chimeric structure, contains a record of the neural bump's trajectory. 

\added{Such a record of history limits the length of trajectories over which the mapping is faithful due to interference with the past trajectory. This restriction is most easily apparent in $2$ dimensions; a self-assembling structure has a ring-like growing front. Growth in a local region of that front by small amounts is still in 1-1 correspondence with bump moving across that growing front. However, interference between larger growth from different parts of the front cannot be captured by the dynamics of the neural bump.}




\section{Conclusions}
We have investigated pattern recognition through self-assembly; the dynamics of self-assembly is able to classify patterns in the concentrations of $N$ molecular species into discrete categories by self-assembling different structures. Even patterns that only modestly resemble an idealized memory are correctly identified and lead to near-exclusive assembly of the corresponding structure.  However, we found that low error rates in classification sets a fundamental lower bound on classification time (i.e., nucleation time) that is independent of physical parameters and only depends on how similar different memories are. Within this bound, we found that classification times can be reduced at the expense of higher error rates by changing parameters like temperature, binding energy and concentrations.

We then related such pattern recognition ability to place cell neural networks. \added{We emphasize that the relationship works on two different levels; (1) the similarity of place field maps and the chemical affinity map shown in Fig.~\ref{fig:Hopfield2Rooms}  (2) a dynamical connection that relies on a quasi-particle approximation of neural activity. Only the low energy states of the neural network can be related to self-assembly as shown in Fig.~\ref{fig:SI_Clump_Tip_correspondence}.} 


The nature of patterns recognized here, based on continuous attractors, is fundamentally different from the original case of point attractors \cite{Hopfield1982-fb}. A $2$-dim memory $m_\alpha$ designed to recognize a given pattern $\vec{c}_a$ will necessarily recognize a further two-parameter family of patterns that correspond to other subsets of the $2$-dim memory. For example, any subset of the image of Leo corresponds to a distinct pattern $\vec{c}$ with perfect overlap $\chi=1$ with the Leo memory and will recall it just as easily. There is no way to avoid this continuous degeneracy and hence no way to create a self-assembly response dedicated to a given pattern $\vec{c}_a$ that is not evoked by any other pattern. (In the neural network, this degeneracy implies that the clump of activity can moved around within a stored environment with no energy cost.)

In contrast, in the original model of Hopfield \cite{Hopfield1982-fb}, each pattern is stored as a discrete point attractor and there is no continuous degeneracy of states. \added{However, the original model of Hopfield was in infinite dimensions; i.e., any neuron could connect to any other, forming a fully connected network.} As shown in \cite{Koyama2001-dc,Derrida1987-ue,Nishimori1995-lm}, short-ranged finite dimensional counterparts of the original model have non-extensive $O(N^{0})$ capacity and cannot store an extensive number of patterns. 

Thus robust pattern recognition with short ranged interactions requires us to go beyond point attractors to continuous attractors. As noted in \cite{Monasson2013-pn} while comparing the phase diagrams of point and continuous attractor models, continuous attractors are more robust to interference between stored patterns than point attractors. Intuitively, storing a single continuous attractor makes a bigger demand of the system than storing a point attractor since a whole set of states needs to be stabilized for the former. Thus, with short ranged interactions, one can still have an extensive capacity when storing continuous attractors \cite{Murugan2015-ps} but not with point attractors \cite{Koyama2001-dc,Derrida1987-ue,Nishimori1995-lm}. 
Consequently, the degeneracy in pattern recognition seen here is intrinsic to self-assembly with short-ranged interactions and cannot be removed through minor modifications of the model. 

We note that it is possible to interpolate between the continuous and point attractor models and thus removing the degeneracy by increasing the dimensionality $d$ of structures. In the limit $d \to \infty$, we assume that each molecule in a structure interacts with all others. Such `mean-field' structures have no geometry and are entirely specified by their composition; e.g., by a Boolean vector $\vec{\sigma}$ of length $N$ where $\sigma_i = 1$ if species $i$ is present in the structure and  $\sigma_i = 0$ otherwise. The energy of the structure is given by $\sum_{ij} J_{ij} \sigma_i \sigma_j$. Thus, the $2^N$ states of such a $d=\infty$ model and their energies are exactly as in the point attractor model of neural networks \cite{Hopfield1982-fb}. For the same reasons, we can store a set of pattern $\sigma_{\mu}$ in the interactions as $J_{ij}=\sum_{\mu} \sigma^i_{\mu} \sigma^j_{\mu}$. If assembly is started with presence of some species $\vec{\sigma}$, the composition of the structure will adjust itself to the closest pre-programmed composition $\vec{\sigma}_\mu$. We also note that the capacity for the self-assembly system \cite{Murugan2015-ps} in $d$-dimensions was found to be $m_c \sim N^{1-1/d}$ which reduces to the classic $m_c \sim N$ point attractor result of \cite{Hopfield1982-fb} as $d \to \infty$. 




Finally, pattern recognition through biophysical dynamics can be exploited in biological and engineering contexts to deploy different self-assembled responses to subtle and hard to define changes in the chemical environment. For example, different transcription factor complexes could be induced by the presence of such imprecise DNA sequence-based or other cues. Similar ideas on pattern recognition of growth factors have been explored in the context of gene regulation during differentiation of stem cells \cite{Lang2014-pz}, protein folding and other contexts \cite{wolynes088,wolynes095,wolynes114,wolynes116,wolynes381}. Our model here is directly implementable with DNA brick assembly \cite{Wei2012-ox} and could be used to recognize patterns as a smart sensor that performs high-dimensional information processing intrinsically, instead of feeding its output to fragile electronics.  Since pattern recognition algorithms and other methods of machine learning often rely on emergent many-body behavior, such artificial intelligence might be realizable through physical and biological dynamics beyond self-assembly.

\section*{Acknowledgements}
We thank Michael Brenner, Nicolas Brunel, John Hopfield, David Huse, Stanislas Leibler, Pankaj Mehta, Remi Monasson, Sidney Nagel,  Sophie Rosay, Zorana Zeravcic and James Zou for discussions. DJS was partially supported by NIH Grant No. K25 GM098875-02.

\appendix

\section{Blurring mask}

We use a blurring mask to easily visualize how closely a concentration pattern $\vec{c}_a$ matches a memory $m_\alpha$. At each pixel (i.e., position) $x$ of a 2-d memory $m_\alpha$, we consider all species in a square box of area $q$ centered at that pixel and find the overlap $\chi(x)$ by finding the fraction of species with high concentration. At position $x$, we then display a gray scale value $(1-\chi(x))*g_{random} + \chi(x)*g_{correct}$, where $g_{correct}$ is the correct gray scale value and  $g_{random}$ is a randomly chosen grayscale value. That is, if every species around $x$ has high concentrations, then $\chi(x) = 1$ and we display the correct pixel. If all species are of low concentrations, then $\chi(x) = 0$ and we display a random gray scale value. Thus, the image displayed is a measure of the contiguity of high concentration species in pattern $\vec{c}_a$ in the spatial arrangement corresponding to memory $m_\alpha$.

In the paper, we use the $\chi$ to compute the rate of nucleation with that region. So $q$ must be larger than the critical nucleation seed at concentrations $c_{low}$ and $c_{high}$. But $q$ must also be small enough, so the $\chi(x)$ computed is representative of the whole region of size $q$. For the images presented here, $q = 8$ was found to satisfy all of these conditions.
With these conditions, the blurring mask provides an immediate visual predictor of nucleation since the blurriness determines the nucleation rate through an exponential function (see Eqn.\ref{eqn:nucleationrate}). If a memory appears coherent over a region, the spontaneous nucleation rate is high in that region and that memory will be rapidly self-assembled. Blurry images have greatly suppressed nucleation rates. 

\section{Gillespie simulation for Self-Assembly}

\added{In our model, particles have distinct binding sites with specific interactions. In the 1D model, each particle has two distinct binding sites, left and right, while in the 2D model, each particle has four distinct sites.} In our simulations, particles cannot ``turn around''; i.e., incoming particles in Fig.~\ref{fig:SI_Clump_Tip_correspondence} always present their left binding site.

We use a Gillespie algorithm to simulate the kinetics shown in Fig.~\ref{fig:SI_Clump_Tip_correspondence} and Fig.\ref{fig:ClassificationSA}; the Gillespie algorithm provides an exact way of measuring physical time using discrete time simulations. At each time step, we consider two kinds of processes: (1) all possible additions of a molecule of any species $i$ to the boundary of the growing structure 
(2) all possible removals of molecules at boundary of the growing structure 
We compute the free energy difference associated with each of these outcomes. For example, if species $i$ is added to the boundary of the structure, the free energy changes by
\begin{equation}
\Delta F = - n E_0 + \mu_{i}
\label{eqn:DeltaFSA}
\end{equation}
where $\mu_i$ is the chemical potential of species $i$ and $n$ is the number of bonds made by species $i$ with its neighbors in the structure (which is, in turn, determined by the interaction matrix $J^{tot}_{ij}$ described in the main text).

We assume that the kinetic rate associated with such a process is $k = e^{-\frac{\Delta F}{2}}$. Removal rates are calculated with the same formula but with the appropriate sign reversals. One of these reactions is randomly chosen and implemented with probability proportional to the corresponding kinetic rate. Physical time is incremented by $t \to t + \frac{1}{\sum_a k^+_a + \sum_x k^-_x}$, in accordance with Gillespie's prescription.

The real time associated with the final frame of simulations in Fig \ref{fig:ClassificationSA} are $\vec{c_1}$: 221221; $\vec{c_2}$: 168449; $\vec{c_3}$: 141202; $\vec{c_4}$: 355121; $\vec{c_5}$: 182573; $\vec{c_6}$: 389395 (all times in arbitrary units). These times are measures of nucleation time in one simulation run from which Fig.\ref{fig:ClassificationSA} was made; growth is much faster than nucleation in our parameter regime.

\added{In this Gillespie simulation, we forbid nucleation at multiple points in the box. Instead we seed the initial structure by one random tile appearing and do not let it disappear. Such a choice does not affect the results and is necessary since our simulations do not include diffusion.  Seeding with a single monomer amounts to following the fate of a generic monomer in the solution. Further, we repeat the simulation and find that the results for a given initial overlap stays the same.}

Finally, Eqn.\ref{eqn:DeltaFSA} can be written in a more intuitive form by defining $f \equiv q E_0  - \mu > 0$, the ``force'' driving self-assembly forward (where $q$ is the range of interactions in the $1$-d model).  If a species binds with energy $E_0$ to only $n = q − w$ of the $q$ molecules in the growing tip in Fig.\ref{fig:SI_Clump_Tip_correspondence}, then Eqn. \ref{eqn:DeltaFSA} reduces to,
\begin{equation}
\Delta F = -f  + w E_0
\label{eqn:DeltaSASimple}
\end{equation}
Thus at low driving forces $f \approx 0$, growth happens near equilibrium since even a molecule making the maximal number $q$ of strong bonds (and thus with $w=0$) binds almost reversibly. At high force $f>0$ (high  concentrations), even species that fail to bind strongly with $w > 0$ particles in the tip can bind. We connect Eqn.\ref{eqn:DeltaSASimple} with corresponding equations for neural networks below.

\section{Neural network simulation}

We perform kinetic Monte Carlo simulations on place cell networks using paired spin flips; at each discrete time step, we randomly pick an `on' neuron $b$ and an `off' neuron $a$ and attempt to flip their states. The free energy change associated with such a move is,
\begin{equation}
\Delta F = -\sum_{c \in clump, c\neq a,b}  (J_{ac} - J_{cb}) 
\label{eqn:DeltaNN}
\end{equation}
We accept such a move with probability $e^{-\Delta F}$.

In our simulations, we drive the clump \cite{Hopfield:2010tk,Ponulak:2013vx} by modifying $J_{ij} \to (1-f/q)J_{ij} + f/q A_{ij}$ by an anti-symmetric component $A_{ij} = - A_{ji}, abs(A_{ij}) = abs(J_{ij})$; this applies a driving force on the clump from left to right in each environment in Fig.~\ref{fig:TrajectoryBranches}. 

We can build some intuition for the driving forces $f$ and connect with self-assembly by re-writing $\Delta F$ for some low energy moves. The left-most neuron in a clump (e.g., neuron $4$ in Fig.~\ref{fig:SI_Clump_Tip_correspondence}) is the easiest to turn off, with a cost $J_{off} = q (E_0 - f/q) = q E_0 - f$. Turning on a neuron $a$ to the right of the clump gives an energy $J_{on} = (q - w)E_0$ where $q-w$ is the number of strong right-ward connections with neurons in the clump made by neuron $a$. Hence the free energy change in moving the clump by turning off the left-most neuron ($4$ in Fig.~\ref{fig:SI_Clump_Tip_correspondence}) and turning on neuron $a$ can be written as,
\begin{equation}
\Delta F_{Clump} = -f + w E_0 
\label{eqn:SimpDeltaNN}
\end{equation}
Such a move is most favorable if $w=0$, i.e., if the new neuron $a$ is fully connected to the clump neurons and thus if the clump moves without breaking up (e.g., neuron $8$ in Fig.~\ref{fig:SI_Clump_Tip_correspondence}). 

Comparing Eqn. \ref{eqn:SimpDeltaNN} and Eqn. \ref{eqn:DeltaSASimple}, we see that low energy transformations of the clump are in correspondence with transformations of the growing structure.

\bibliographystyle{unsrt}
\bibliography{OldRefs_Papers2_Bibtex,Paperpile_-_Dec_15_BibTeX_Export,wolynes}

\begin{thebibliography}{10}

\bibitem{Graves2013-qs}
A~Graves, A~r.~Mohamed, and G~Hinton.
\newblock Speech recognition with deep recurrent neural networks.
\newblock In {\em 2013 {IEEE} International Conference on Acoustics, Speech and
  Signal Processing}, pages 6645--6649, May 2013.

\bibitem{Krizhevsky2012-jc}
Alex Krizhevsky, Ilya Sutskever, and Geoffrey~E Hinton.
\newblock {ImageNet} classification with deep convolutional neural networks.
\newblock In F~Pereira, C~J~C Burges, L~Bottou, and K~Q Weinberger, editors,
  {\em Advances in Neural Information Processing Systems 25}, pages 1097--1105.
  Curran Associates, Inc., 2012.

\bibitem{Hopfield1982-fb}
J~J Hopfield.
\newblock Neural networks and physical systems with emergent collective
  computational abilities.
\newblock In {\em Proceedings of the International Association for Shell and
  Spatial Structures ({IASS}) Symposium 2009}, 1~January 1982.

\bibitem{Purvis2013-io}
Jeremy~E Purvis and Galit Lahav.
\newblock Encoding and decoding cellular information through signaling
  dynamics.
\newblock {\em Cell}, 152(5):945--956, 28~February 2013.

\bibitem{Levine2013-yd}
J~H Levine, Y~Lin, and M~B Elowitz.
\newblock Functional roles of pulsing in genetic circuits.
\newblock {\em Science}, 342(6163):1193--1200, 5~December 2013.

\bibitem{Brubaker2015-ft}
Sky~W Brubaker, Kevin~S Bonham, Ivan Zanoni, and Jonathan~C Kagan.
\newblock Innate immune pattern recognition: a cell biological perspective.
\newblock {\em Annu. Rev. Immunol.}, 33:257--290, 2~January 2015.

\bibitem{Murugan2015-ps}
Arvind Murugan, Zorana Zeravcic, Michael~P Brenner, and Stanislas Leibler.
\newblock Multifarious assembly mixtures: Systems allowing retrieval of diverse
  stored structures.
\newblock {\em Proceedings of the National Academy of Sciences}, 112(1):54--59,
  6~January 2015.

\bibitem{Amit:1985tj}
Daniel Amit, Hanoch Gutfreund, and H~Sompolinsky.
\newblock {Storing infinite numbers of patterns in a spin-glass model of neural
  networks}.
\newblock {\em Phys. Rev. Lett.}, 55(14):1530--1533, September 1985.

\bibitem{Hertz:1991ud}
John Hertz, Anders Krogh, and Richard Palmer.
\newblock {\em {Introduction to the Theory of Neural Computation}}.
\newblock Basic Books, January 1991.

\bibitem{Amit1985-ls}
Daniel~J Amit, Hanoch Gutfreund, and Haim Sompolinsky.
\newblock Spin-glass models of neural networks.
\newblock {\em Phys. Rev. A}, 32(2):1007, 1~January 1985.

\bibitem{MacKay2003-nq}
David J~C MacKay.
\newblock {\em Information Theory, Inference and Learning Algorithms}.
\newblock Cambridge University Press, 25~September 2003.

\bibitem{Burak2012-bu}
Yoram Burak and Ila~R Fiete.
\newblock Fundamental limits on persistent activity in networks of noisy
  neurons.
\newblock {\em Proc. Natl. Acad. Sci. U. S. A.}, 109(43):17645--17650,
  23~October 2012.

\bibitem{Chaudhuri2016-lh}
Rishidev Chaudhuri and Ila Fiete.
\newblock Computational principles of memory.
\newblock {\em Nat. Neurosci.}, 19(3):394--403, 23~February 2016.

\bibitem{Seung1997-ly}
H~Sebastian Seung.
\newblock Learning continuous attractors in recurrent networks.
\newblock In {\em {NIPS}}, volume~97, pages 654--660, 1997.

\bibitem{Wu2008-iw}
Si~Wu, Kosuke Hamaguchi, and Shun-Ichi Amari.
\newblock Dynamics and computation of continuous attractors.
\newblock {\em Neural Comput.}, 20(4):994--1025, April 2008.

\bibitem{Monasson2013-pn}
R{\'e}mi Monasson and Sophie Rosay.
\newblock Crosstalk and transitions between multiple spatial maps in an
  attractor neural network model of the hippocampus: Phase diagram.
\newblock {\em Physical review E}, 87(6):062813, 1~January 2013.

\bibitem{Monasson2014-nu}
R~Monasson and S~Rosay.
\newblock Crosstalk and transitions between multiple spatial maps in an
  attractor neural network model of the hippocampus: Collective motion of the
  activity.
\newblock {\em Physical review E}, 89(3), 1~January 2014.

\bibitem{Battaglia:1998bm}
F~Battaglia and A~Treves.
\newblock {Attractor neural networks storing multiple space representations: A
  model for hippocampal place fields}.
\newblock {\em Phys. Rev. E}, 58(6):7738--7753, December 1998.

\bibitem{Seung2000-bk}
H~S Seung, D~D Lee, B~Y Reis, and D~W Tank.
\newblock Stability of the memory of eye position in a recurrent network of
  conductance-based model neurons.
\newblock {\em Neuron}, 26(1):259--271, April 2000.

\bibitem{Hopfield:2010tk}
John~J Hopfield.
\newblock {Neurodynamics of mental exploration}.
\newblock {\em Proceedings of the National Academy of Sciences of the United
  States of America}, 107(4):1648--1653, 2010.

\bibitem{Hopfield2015-wt}
John~J Hopfield.
\newblock Understanding emergent dynamics: Using a collective activity
  coordinate of a neural network to recognize {Time-Varying} patterns.
\newblock {\em Neural Comput.}, 27(10):2011--2038, 1~October 2015.

\bibitem{Fink:2001ki}
Thomas Fink and Robin Ball.
\newblock {How Many Conformations Can a Protein Remember?}
\newblock {\em Phys. Rev. Lett.}, 87(19):198103, October 2001.

\bibitem{Barish:2009te}
Robert~D Barish, Rebecca Schulman, Paul~WK Rothemund, and Erik Winfree.
\newblock {An information-bearing seed for nucleating algorithmic
  self-assembly}.
\newblock {\em Proceedings of the National Academy of Sciences of the United
  States of America}, 106(15):6054--6059, 2009.

\bibitem{wolynes088}
Mark~S Friedrichs and Peter~G Wolynes.
\newblock Toward protein tertiary structure recognition by means of associative
  memory hamiltonians.
\newblock {\em Science}, 246(4928):371, 1989.

\bibitem{wolynes095}
M~Sasai and PG~Wolynes.
\newblock Molecular theory of associative memory hamiltonian models of protein
  folding.
\newblock {\em Physical review letters}, 65(21):2740, 1990.

\bibitem{wolynes114}
M~Sasai and PG~Wolynes.
\newblock Unified theory of collapse, folding, and glass transitions in
  associative-memory hamiltonian models of proteins.
\newblock {\em Physical review A}, 46(12):7979, 1992.

\bibitem{wolynes116}
Henrik~G Bohr and Peter~G Wolynes.
\newblock Initial events of protein folding from an information-processing
  viewpoint.
\newblock {\em Physical Review A}, 46(8):5242, 1992.

\bibitem{wolynes381}
Nicholas~P Schafer, Bobby~L Kim, Weihua Zheng, and Peter~G Wolynes.
\newblock Learning to fold proteins using energy landscape theory.
\newblock {\em Israel journal of chemistry}, 54(8-9):1311--1337, 2014.

\bibitem{Ke2012-lm}
Y~Ke, L~L Ong, W~M Shih, and Peng Yin.
\newblock {Three-Dimensional} structures {Self-Assembled} from {DNA} bricks.
\newblock {\em Science}, 338(6111):1177--1183, 29~November 2012.

\bibitem{Wei2012-ox}
Bryan Wei, Mingjie Dai, and Peng Yin.
\newblock Complex shapes self-assembled from single-stranded {DNA} tiles.
\newblock {\em Nature}, 485(7400):623--626, 30~May 2012.

\bibitem{Colgin2010-rd}
Laura~L Colgin, Stefan Leutgeb, Karel Jezek, Jill~K Leutgeb, Edvard~I Moser,
  Bruce~L McNaughton, and May-Britt Moser.
\newblock Attractor-map versus autoassociation based attractor dynamics in the
  hippocampal network.
\newblock {\em J. Neurophysiol.}, 104(1):35--50, July 2010.

\bibitem{Jezek2011-sz}
Karel Jezek, Espen~J Henriksen, Alessandro Treves, Edvard~I Moser, and
  May-Britt Moser.
\newblock Theta-paced flickering between place-cell maps in the hippocampus.
\newblock {\em Nature}, 478(7368):246--249, 28~September 2011.

\bibitem{Wills2005-hz}
Tom~J Wills, Colin Lever, Francesca Cacucci, Neil Burgess, and John O'Keefe.
\newblock Attractor dynamics in the hippocampal representation of the local
  environment.
\newblock {\em Science}, 308(5723):873--876, 6~May 2005.

\bibitem{Kubie:1991vq}
John~L Kubie and Robert~U Muller.
\newblock {Multiple representations in the hippocampus}.
\newblock {\em Hippocampus}, 1(3):240--242, 1991.

\bibitem{Curto:2008vr}
Carina Curto and Vladimir Itskov.
\newblock {Cell groups reveal structure of stimulus space}.
\newblock {\em PLoS Comp Biol}, 4(10):e1000205, 2008.

\bibitem{Pfeiffer2013-qn}
Brad~E Pfeiffer and David~J Foster.
\newblock Hippocampal place-cell sequences depict future paths to remembered
  goals.
\newblock {\em Nature}, 497(7447):74--79, 2~May 2013.

\bibitem{Ponulak2013-op}
Filip Ponulak and John~J Hopfield.
\newblock Rapid, parallel path planning by propagating wavefronts of spiking
  neural activity.
\newblock {\em Front. Comput. Neurosci.}, 7, 1~January 2013.

\bibitem{Wu2005-sw}
Si~Wu and Shun-Ichi Amari.
\newblock Computing with continuous attractors: stability and online aspects.
\newblock {\em Neural Comput.}, 17(10):2215--2239, October 2005.

\bibitem{Jezek:2011ib}
Karel Jezek, Espen~J Henriksen, Alessandro Treves, Edvard~I Moser, and
  May-Britt Moser.
\newblock {Theta-paced flickering between place-cell maps in the hippocampus}.
\newblock {\em Nature}, 478(7368):246--249, September 2011.

\bibitem{Hedges2014-af}
Lester~O Hedges, Ranjan~V Mannige, and Stephen Whitelam.
\newblock Growth of equilibrium structures built from a large number of
  distinct component types.
\newblock {\em Soft Matter}, 10(34):6404--6416, 1~January 2014.

\bibitem{Murugan2015-wb}
Arvind Murugan, James Zou, and Michael~P Brenner.
\newblock Undesired usage and the robust self-assembly of heterogeneous
  structures.
\newblock {\em Nat. Commun.}, 6:6203, 11~February 2015.

\bibitem{Jacobs2013-rk}
W~M Jacobs and D~Frenkel.
\newblock Predicting phase behavior in multicomponent mixtures.
\newblock {\em J. Chem. Phys.}, 1~January 2013.

\bibitem{Jacobs2015-jt}
William~M Jacobs, Aleks Reinhardt, and Daan Frenkel.
\newblock Communication: Theoretical prediction of free-energy landscapes for
  complex self-assembly.
\newblock {\em J. Chem. Phys.}, 142(2):021101, 14~January 2015.

\bibitem{Haxton2013-bm}
Thomas~K Haxton and Stephen Whitelam.
\newblock Do hierarchical structures assemble best via hierarchical pathways?
\newblock {\em Soft Matter}, 9(29):6851--6861, 1~January 2013.

\bibitem{Whitelam2012-nt}
Stephen Whitelam, Rebecca Schulman, and Lester Hedges.
\newblock {Self-Assembly} of multicomponent structures in and out of
  equilibrium.
\newblock {\em Phys. Rev. Lett.}, 109(26):265506, 1~December 2012.

\bibitem{Levy:2006wv}
Emmanuel~D Levy, Jose~B Pereira-Leal, Cyrus Chothia, and Sarah~A Teichmann.
\newblock {3D complex: a structural classification of protein complexes}.
\newblock {\em PLoS Comp Biol}, 2(11):e155, 2006.

\bibitem{Koyama2001-dc}
Shinsuke Koyama.
\newblock Storage capacity of two-dimensional neural networks.
\newblock {\em Physical review E}, 65(1):016124, 1~January 2001.

\bibitem{Derrida1987-ue}
B~Derrida, E~Gardner, and A~Zippelius.
\newblock An exactly solvable asymmetric neural network model.
\newblock {\em EPL}, 4(2):167, 1987.

\bibitem{Nishimori1995-lm}
H~Nishimori, W~Whyte, and D~Sherrington.
\newblock Finite-dimensional neural networks storing structured patterns.
\newblock {\em Phys. Rev. E Stat. Phys. Plasmas Fluids Relat. Interdiscip.
  Topics}, 51(4):3628--3642, April 1995.

\bibitem{Lang2014-pz}
Alex~H Lang, Hu~Li, James~J Collins, and Pankaj Mehta.
\newblock Epigenetic landscapes explain partially reprogrammed cells and
  identify key reprogramming genes.
\newblock {\em PLoS Comput. Biol.}, 10(8):e1003734, 1~January 2014.

\bibitem{Ponulak:2013vx}
Filip Ponulak and John~J Hopfield.
\newblock {Rapid, parallel path planning by propagating wavefronts of spiking
  neural activity}.
\newblock {\em Frontiers in computational neuroscience}, 7, 2013.

\end{thebibliography}

\end{document}